\documentclass[a4paper, 11pt]{article}
\usepackage[english]{babel}
\makeatletter

\usepackage[top=2cm, bottom=2cm, inner=1.5cm, outer=1.5cm]{geometry}
\usepackage{graphicx, color}
\usepackage{amsfonts, amsmath, amssymb, slashed,dsfont}
\usepackage{tensor}
\usepackage[pdftex]{hyperref}
\usepackage{epsfig}
\usepackage{pifont}
\usepackage{enumitem}
\usepackage{bm}
\usepackage{mathrsfs} % Allows \mathscr
\usepackage{amsmath,amssymb,slashed}
\usepackage[normalem]{ulem} % Allows to strickout
\usepackage{tikz}
\usetikzlibrary{positioning}
\usetikzlibrary{decorations.markings}
\usetikzlibrary{arrows.meta}

\numberwithin{equation}{section}

\newcommand{\tops}[2]{\texorpdfstring{#1}{#2}}

\newcommand\algA{{\mathbb{A}}}
\newcommand\lHm{\td \mu}
\newcommand\rHm{\td \mu_R}
\newcommand{\td}{\textnormal{d}}
\newcommand{\tdl}[2]{\textnormal{d}^{#1}{#2}\;}

\begin{document}

\begin{center}
\begin{minipage}{.9\textwidth}
	\begin{center}
	\begin{huge}
		On the UV/IR mixing of Lie algebra-type noncommutatitive $\phi^4$-theories
	\end{huge}
	
	\vspace{20pt}
	{\Large Kilian Hersent}%
	\footnote{
		IJCLab, Universit\'{e} Paris-Saclay, CNRS/IN2P3, 91405 Orsay, France. \href{mailto:kilian.hersent@universite-paris-saclay.fr}{kilian.hersent@universite-paris-saclay.fr}
	}

	\vspace{40pt}
	{\large \textbf{Abstract}}
	\end{center}
	
	We show that a UV divergence of the propagator integral implies the divergences of the UV/IR mixing in the two-point function at one-loop for a $\phi^4$-theory on a generic Lie algebra-type noncommutative space-time. The UV/IR mixing is defined as a UV divergence of the planar contribution and an IR singularity of the non-planar contribution, the latter being due to the former UV divergence, and the UV finiteness of the non-planar contribution. Some properties of this general treatment are discussed. The UV finiteness of the non-planar contribution and the renormalizability of the theory are not treated but commented. Applications are performed for the Moyal space, having a UV/IR mixing, and the $\kappa$-Minkowski space for which the two-point function at one-loop is finite.
\end{minipage}
\end{center}

\vspace{20pt}
\paragraph{}
Noncommutative space-times correspond to deformations of usual space-times through the introduction of a noncommutative product, usually called the star-product. This deformation causes the coordinate functions not to commute anymore, thus generating a fuzziness in the space-time \cite{DFR}. The notion of point is thus blurred, a behaviour expected to happened in a quantum space-time. Therefore, the noncommutative space-times are expected to account for some quantum gravity effects, at least in some regime.

Though, the noncommutative spaces were first studied to cope with ultraviolet (UV) divergences in field theories \cite{Snyder_1947}. Field theories on noncommutative spaces were heavily studied \cite{Hersent_2022} and have shown a pathological behaviour called the UV/IR mixing.

\paragraph{}
The UV/IR mixing takes its root in the fact that the $2$-point function, at one-loop order, now splits into two contributions : the planar contribution and the non-planar one. On the one hand, the planar contribution triggers a UV divergence, similar to the commutative one. On the other hand, the non-planar contribution is usually UV finite but infra-red (IR) singular. It is further called mixing since the two divergences come from the same source. The mixing is often considered to be a plague for the noncommutative field theories since it would trigger non-renormalisability.

This phenomenon was first evidenced on the Moyal space in the context of scalar field theories \cite{Minwalla_2000, Micu_2001} and shortly extended to noncommutative gauge theories \cite{Matusis_2000}. For reviews on the Moyal space see e.g.\ \cite{Blaschke_2010}. Still, it was also evidenced for scalar theories on other noncommutative space-times like the $\kappa$-Minkowski space \cite{Grosse_2006, Poulain_2019}, the $\rho$-Minkowski space \cite{Dimitrijevic_2018, Hersent_2023}, or others \cite{Robbins_2003}. The mixing of noncommutative gauge theories on other spaces then Moyal is still poorly studied \cite{Hersent_2022}.

\paragraph{}
There has been solution proposals on the Moyal space, notably the Grosse-Wulkenhaar model \cite{Grosse_2005a, Grosse_2003, Grosse_2005b} and the IR damping model \cite{Gurau_2009}, among others \cite{Mirza_2006, Jafari_Salim_2006, Schenkel_2010}. Still, none of these solutions try to analyse the roots of the mixing or extend to other noncommutative space-times then Moyal.

In this paper, we build a general approach to field theory based on the momentum formalism, always present in Lie algebra-type noncommutative spaces. We prove that the divergence of the integral of the propagator is the source of the mixing of the $2$-point function at one-loop.

\paragraph{}
In order to be accurate, let us first define what we mean about ``UV/IR mixing''. We define the UV/IR mixing at one-loop as having three components:
\begin{enumerate}[label= (\roman*)]
	\item \label{it:UV}
	The planar contribution is diverging in the UV,
	\item \label{it:IR}
	The non-planar contribution is singular in the IR, this IR singularity being due to the UV divergence of \ref{it:UV}.
	\item \label{it:fin}
	The non-planar contribution is UV finite.
\end{enumerate}
This definition is taken to be broad and avoids the notion of renormalisability on purpose, as discussed in section \ref{subsec:UV_IR_div}. Furthermore, this definition does not take into account some physical intuition the mixing could have, like the fact that it can be linked to the existence of string modes which are simultaneously short-ranged (UV) and long-ranged (IR) \cite{Steinacker_2016}.

We here show that the divergence of $\int \lHm(k) K^{-1}(k)$ is equivalent to \ref{it:UV} and \ref{it:IR}, where $\lHm$ is the left Haar measure of the momentum Lie group and $K^{-1}$ is the propagator. The requirement \ref{it:fin} is necessary to distinguish between the commutative and noncommutative cases, still its fate is not clear in the general case.

Finally, we apply this formalism to two examples. The first one is the Moyal space for which the chosen propagator triggers a mixing. The second one is the $\kappa$-Minkowski space for which the propagator avoids the mixing, even making the $2$-point function finite at one-loop. The momentum ordering freedom is discussed.

\paragraph{}
The framework is set-up in section \ref{sec:set-up}. Some notions of integration on groups are gathered in section \ref{sec:gp}, in order to perform the analysis of the $\phi^4$-theory detailed in section \ref{sec:sft}. The application of the previous formalism is done in section \ref{sec:eg}. Finally, appendix \ref{apxsec:form} regroups some useful formulas and the extended computations are performed in appendix \ref{apxsec:comp}.

\section{From noncommutativity of space to deformed composition of momenta}
\label{sec:set-up}
\paragraph{}
We consider here the noncommutative space to be a $*$-algebra $\algA = \big(C^\infty(\mathcal{M}), \star \big)$, where $\mathcal{M}$ is a $d+1$-dimensional space-time, equipped with a noncommutative product $\star$ and an involution ${}^\dagger$. We also assume that one can define global coordinate%
\footnote{
	One could equivalently consider that those coordinates corresponds to local coordinates on a patch of open set of $\mathcal{M}$.	
}
functions $\{x^\mu\}_{\mu = 0, \dots, d}$ on $\mathcal{M}$ and that they satisfy a ``Lie algebra type noncommutativitity'' for the star-product. Explicitly, the coordinates forms a Lie algebra $\mathfrak{g}$ with bracket
\begin{align}
	[x^\mu, x^\nu]_\star 
	= x^\mu \star x^\nu - x^\nu \star x^\mu
	= \tensor{f}{^{\mu\nu}_\rho} x^\rho,
	\label{eq:coord_nc}
\end{align}
where $\tensor{f}{^{\mu\nu}_\rho} \in \mathbb{C}$ is called the structure constant and the Einstein summation convention is applied.

\paragraph{}
When going to momentum space, by considering the Fourier transform, the noncommutative star-product $\star$ turns into a noncommutative composition law for momentum, denoted $\oplus$. This is always possible thanks to the Baker-Campbell-Hausdorff formula. Explicitly, one can write
\begin{align}
	e^{i p_\mu x^\mu} \star e^{i q_\mu x^\mu}
	= e^{i (p_\mu \oplus q_\mu) x^\mu}
	\label{eq:oplus_def}
\end{align}
where $p_\mu$ and $q_\mu$ are momenta. These momenta are coordinates on the Lie group $(\mathcal{G}, \oplus)$, which is in correspondence with the Lie algebra $\mathfrak{g}$. We assume that $\mathcal{G}$ is locally compact so that a left Haar measure $\lHm$ exists, as detailed in section \ref{sec:gp}.

One can define an inverse to $\oplus$ thanks to the involution ${}^\dagger$, through
\begin{align}
	\left( e^{i p_\mu x^\mu} \right)^\dagger
	= e^{i (\ominus p_\mu) x^\mu}.
	\label{eq:ominus_def}
\end{align}
Then, one can show that the following rules stands
\begin{subequations}
	\label{eq:op_prop}
\begin{align}
	p_\mu \oplus (\ominus p_\mu) &= 0,
	\label{eq:op_prop_inv}\\
	\ominus ( \ominus p_\mu) &= p_\mu,
	\label{eq:op_prop_id} \\
	\ominus(p_\mu \oplus q_\mu)
	&= (\ominus q_\mu) \oplus (\ominus p_\mu).
	\label{eq:op_prop_dist}
\end{align}
\end{subequations}
The property \eqref{eq:op_prop_dist} might be puzzling at first since $\oplus$ is denoted as an additive (abelian) law. However, it behaves like a multiplicative law since it is noncommutative. In the following, we will write for simplicity $p_\mu \ominus q_\mu = p_\mu \oplus (\ominus q_\mu)$.

\paragraph{}
At this point several remarks are in order.

\paragraph{}
First, the commutative limit of the $\oplus$ law is straightforwardly the usual $+$ law. Indeed, the proper commutative limit is to consider that the star-product $\star$ is the point-wise product. In this case, the exponential product \eqref{eq:oplus_def} gives $\oplus = +$. Similarly, the involution ${}^\dagger$ becomes the complex conjugation so that \eqref{eq:ominus_def} implies $\ominus = -$.

\paragraph{}
The laws $\oplus$ and $\ominus$ can be expressed thanks to the structure constant $\tensor{f}{^{\mu\nu}_\rho}$ of \eqref{eq:coord_nc}, through the Baker-Campbell-Hausdorff formula. In practice, this formula is very involved and can only be made explicit in specific cases (see \cite{Van_Brunt_2015, Van_Brunt_2018} for the general theory, and \cite{Poulain_2019, Hersent_2023} for examples). Therefore, we will only work with the $\oplus$ and $\ominus$ laws in the following.

\paragraph{}
As pointed out in \cite{Mercati_2018}, there is an ordering ambiguity for expressing the exponential $e^{i p_\mu x^\mu}$ in \eqref{eq:oplus_def}, since all expressions such as
\begin{align}
	e^{i p_\mu x^\mu}
	&= e^{i (p_0 x^0 + \dots + p_d x^d)}, &
	e^{i p_\mu x^\mu}
	&= e^{i p_0 x^0} \cdots e^{i p_d x^d}, &
	e^{i p_\mu x^\mu}
	&= e^{i p_d x^d} \cdots e^{i p_0 x^0},
	\label{eq:mom_order}
\end{align}
are the same in the commutative limit. Thus, this ordering prescription is a specificity to the noncommutative case and these ordering generates different laws $\oplus$.

Still, in the case of $\kappa$-Minkowski, they are all equal up to momentum change of coordinate \cite{Mercati_2018}. Therefore, if we consider a momentum diffeomorphic invariant theory, then the ordering \eqref{eq:mom_order} does not matter.

In the following, the ordering will only be discussed in examples (see section \ref{sec:eg}), since the $\oplus$ law remains general in the core of the paper.

\paragraph{}
Some authors showed that the translation invariance of the star-product in a $\phi^4$ theory described leads necessarily to UV/IR mixing \cite{Galluccio_2009}. If we denote by $\mathcal{T}_z(f)(x) = f(x + z)$ the translation operator, then a translation invariant star-product satisfies
\begin{align}
	\mathcal{T}_z(f \star g)
	&= \mathcal{T}_z (f) \star \mathcal{T}_z (g).
\end{align}
Applying this relation in the case of plane waves, one gets that such a star-product must satisfy
\begin{align*}
	\mathcal{T}_z \left( (e^{i p_\mu \cdot^\mu} \star e^{i q_\mu \cdot^\mu})(x) \right)
	&= e^{i(p_\mu \oplus q_\mu) (x^\mu + z^\mu)} \\
	= \left(\mathcal{T}_z (e^{i p_\mu \cdot^\mu}) \star \mathcal{T}_z (e^{i q_\mu \cdot^\mu}) \right)(x)
	&= e^{i (p_\mu \oplus q_\mu) x^\mu} e^{i (p_\mu + q_\mu) z^\mu},
\end{align*}
and so taking $z^\mu = (0, \dots, 0, \overset{\mu}{1}, 0, \dots, 0)$, this amounts to
\begin{align}
	p \oplus q = p + q.
\end{align}
Therefore, in the case of a Lie algebra-type noncommutativity, translation invariance of the star-product implies that the addition law of momenta is undeformed, which only happen in the case of Moyal \eqref{eq:Moy_Lie_gp}, where the deformation is contained in the unit sector as a phase term.

Moreover, one should stress that the analysis of \cite{Galluccio_2009} was only carried out for action of the form \eqref{eq:action} since the counter-term of \cite{Gurau_2009} is translation invariant.

\section{Integration on groups}
\label{sec:gp}
\paragraph{}
We here recall some elements of integration on locally compact groups. For a more complete treatment of harmonic analysis one can refer to textbooks like \cite{Deitmar_2014}. 

\paragraph{}
In order to define quantity such as an inverse Fourier transform, one needs a way to sum the momentum space and so a measure. In the commutative case, such a space corresponds to $\mathbb{R}^{d+1}$ so that the usual Lebesgue measure can be used. In the present case, the space of momentum is a Lie group $\mathcal{G}$, hypothesised to be locally compact. In such a case, two measures always exists which are the unique left (noted $\lHm$) and the unique right (noted $\rHm$) invariant Haar measures. Their invariance property are defined by
\begin{align*}
	\lHm( p \oplus q )
	&= \lHm(q), &
	\rHm( p \oplus q )
	&= \rHm(p).
\end{align*}
Their uniqueness property states that every left invariant measure corresponds to $\lHm$ up to a positive factor, and similarly for the right invariance.

Now, for $p \in \mathcal{G}$, we consider the measure $\lHm_p(U) = \lHm(U \oplus p)$, for any $U \subset \mathcal{G}$. Then, one has for any $q \in \mathcal{G}$,
\begin{align*}
	\lHm_p(q \oplus U)
	= \lHm( q \oplus U \oplus p)
	= \lHm (U \oplus p)
	= \lHm_p(U),
\end{align*}
using the left invariance of $\lHm$. Therefore, $\lHm_p$ is left invariant. By uniqueness, this implies that it exists $\Delta(p) > 0$ such that $\lHm_p = \Delta(p) \lHm$. The function we just constructed $\Delta : \mathcal{G} \to \mathbb{R}_+\backslash\{0\}$ is called the modular function. This modular function is a homomorphism, meaning 
\begin{align*}
	\Delta(p \oplus q) = \Delta(p) \Delta(q), &&
	\Delta(\ominus p) = \Delta(p)^{-1}, &&
	\Delta(0) = 1. 
\end{align*}
One can also show that $\Delta$ is continuous. From its construction, the modular function can be used to change variable in the integrand. Explicitly, considering the left invariant Haar measure $\lHm$, 
\begin{subequations}
	\label{eq:Haar_meas}
\begin{align}
	\int \lHm(p)\ f(q \oplus p)
	& \overset{p \to \ominus q \oplus p}{=}
	\int \lHm(p)\ f(p),
	\label{eq:Haar_meas_inv} \\
	\int \lHm(p)\ f(p \oplus q)
	& \overset{p \to p \ominus q}{=}
	\int \lHm(p)\ \Delta(\ominus q) f(p).
	\label{eq:Haar_meas_uni}
\end{align}
\end{subequations}
The equation \eqref{eq:Haar_meas_inv} comes from the fact that $\lHm$ is left invariant and the equation \eqref{eq:Haar_meas_uni} comes from the modular function construction above.

\paragraph{}
The modular function ``quantifies'' the difference between the left and the right Haar measures. Indeed, if one consider $\lHm_\ominus(p) = \lHm (\ominus p)$, then for any $q \in \mathcal{G}$, one has
\begin{align*}
	\lHm_\ominus (q \oplus p)
	= \lHm (\ominus( q \oplus p) )
	= \lHm( \ominus p \ominus q)
	= \lHm (\ominus q)
	= \lHm_\ominus (q),
\end{align*}
so that $\lHm_\ominus$ is right-invariant. By uniqueness, $\lHm_\ominus = \rHm$, up to an irrelevant scaling factor, and one has as a property that $\lHm (\ominus p) = \Delta(\ominus p)\ \lHm(p)$, so that 
\begin{align}
	\rHm(p) = \Delta(\ominus p)\ \lHm(p).
	\label{eq:Haarm_l_vs_r}
\end{align}

In the case where $\Delta = 1$, the group $\mathcal{G}$ is said to be unimodular, and the right and left Haar measures are equal. In the following, the group $\mathcal{G}$ will not be necessarily unimodular.

\paragraph{}
We can now integrate quantities on $\mathcal{G}$, but one notion remain to be discussed for our purpose : the Dirac delta function. Even if the delta function is not necessarily defined in this context, we will introduce it in the physics way. However, there is a need of consistency with respect to the considered measure. Indeed, if we use the left invariant Haar measure then the following identities are needed
\begin{subequations}
	\label{eq:delta_rule}
\begin{align}
	\int \lHm(p)\ f(p) \delta(p \oplus q)
	&= \Delta( \ominus q) f(\ominus q),
	\label{eq:delta_left}\\
	\int \lHm(p)\ f(p) \delta(q \oplus p)
	&= f(\ominus q),
	\label{eq:delta_right}
\end{align}
\end{subequations}
for any function $f$. These rules solve a possible mismatch when computing $\int \lHm(p)\ f(p) \delta(p \oplus q)$. Performing the delta integration through the usual physicist way (that is imposing $p = \ominus q$), or performing the change of variable $p \to p \ominus q$ and then integrating the $\delta$, would give two different results otherwise.

The previous rule can be summarized in a deformed cyclic rule of the $\delta$ function given by
\begin{align}
	\delta( p \oplus q )
	= \Delta(\ominus q)\ \delta( q \oplus p )
	\label{eq:delta_cycl}
\end{align}

One also deduces from \eqref{eq:delta_rule} that
\begin{align}
	\int \lHm(p)\ f(p) \delta(p \ominus q)
	&= \int \lHm(p)\ f(p) \delta(q \ominus p)
	= \Delta(q) f(q),
	\label{eq:delta_minus} \\
	\delta(\ominus p) &= \delta(p).
	\label{eq:minus_delta}
\end{align}
We now have all the tool we need to build the field theory. Note that the following analysis is carried out with the left Haar measure. Still, considering the right invariant Haar measure would lead to exactly the same results. One should note that the relations \eqref{eq:delta_rule} to \eqref{eq:minus_delta} changes when considering the right Haar measure.

\section{Uncharged scalar field theory}
\label{sec:sft}
\paragraph{}
We here consider a $\phi^4$ theory with the following action
\begin{align}
	S(\phi)
	&= \int \tdl{d+1}{x} \partial_\mu \phi \star \partial^\mu \phi(x)
	+ m^2 \phi \star \phi(x)
	+ \frac{\lambda}{4!}\ \phi \star \phi \star \phi \star \phi (x),
	\label{eq:action}
\end{align}
where $m$ is the mass of the field, $\lambda$ is the coupling constant and the ``background metric'' $\mathrm{g}$, in $\partial_\mu \partial^\mu = \mathrm{g}^{\mu\nu} \partial_\mu \partial_\nu$, is the metric of $\mathcal{M}$ and thus will remain general, but symmetric. We consider here a trace $\int \tdl{d+1}{x}$ over $\mathbb{A}$, possibly inherited from an integral over $\mathcal{M}$. Note that this trace may be non-cyclic as we shall see in section \ref{subsec:trace_commu}.

\subsection{Deformed Fourier transform}
\paragraph{}
Going into momentum space, one consider the Fourier transform of $\phi$, also denoted $\phi$, defined by
\begin{align}
	\phi(p) = \int \tdl{d+1}{x} e^{i (\ominus p_\mu) x^\mu} \star \phi(x), &&
	\phi(x) = \frac{1}{(2\pi)^{d+1}} \int \lHm(p)\ \phi(p) e^{i p_\mu x^\mu}.
	\label{eq:F_trans_phi}
\end{align}
One should note that the exponential is $\star$-multiplied on the left in the first term of \eqref{eq:F_trans_phi} to match the fact that we used the left invariant Haar measure $\lHm$ in the inverse Fourier transform. The right invariant Haar measure would have needed a right $\star$-multiplication.

Rewriting the action \eqref{eq:action} with the Fourier transform \eqref{eq:F_trans_phi}, one obtains
\begin{align}
\begin{aligned}
	S(\phi)
	=&\ \int \lHm(k)\ \lHm(\tilde{k})\ 
		\frac{1}{2} \left(k_\mu \tilde{k}^\mu + m^2\right) \delta(k \oplus \tilde{k}) \phi(k) \phi(\tilde{k}) \\
	&+ \frac{\lambda}{4!} \int \lHm(k^1)\ \lHm(k^2)\ \lHm(k^3)\ \lHm(k^4)\ 
		\delta \Big(k^1 \oplus k^2 \oplus k^3 \oplus k^4 \Big)\  \phi(k^1) \phi(k^2) \phi(k^3) \phi(k^4).
\end{aligned}
	\label{eq:action_mom}
\end{align}
For simplicity, we write $K(k) = \frac{1}{2} \big(k^\mu (\ominus k_\mu) + m^2 \big)$ and $V(k^1 k^2 k^3 k^4) = \delta(k^1 \oplus k^2 \oplus k^3 \oplus k^4)$. Therefore, the previous action writes
\begin{align*}
	S(\phi)
	&= \frac{1}{2} \int \lHm(k)\  \phi(k) K(k) \phi(\ominus k)
	 + \frac{\lambda}{4!} \int \lHm(k^j)\  V(k^1 k^2 k^3 k^4)\ \phi(k^1) \phi(k^2) \phi(k^3) \phi(k^4) \\
	&= S_0(\phi) + S_{\text{int}}(\phi).
\end{align*}

\paragraph{}
Some authors \cite{Fatollahi_2007, Komaie_Moghaddam_2007, Komaie_Moghaddam_2008} have considered a similar formalism based on the Lie group of momentum space to study $\phi^4$ theory on $\mathbb{R}^3_\lambda$.

\subsection{Perturbative expension at one-loop}
\paragraph{}
At this point, we perform the textbook construction of the perturbative expansion of the generating functional of the connected Green functions. Explicitly, we consider the generating function
\begin{align}
	\mathcal{Z}(J)
	&= \int \tdl{}{\phi}\ \mathrm{exp} \left( -S(\phi) + \frac{1}{2} \int J \star \phi + \frac{1}{2} \int \phi \star J \right)
	\label{eq:gen_func}
\end{align}
where $J$ is a source term and we noted $\int \phi \star J = \int \tdl{d+1}{x} \phi \star J(x)$. The symmetrised star-product $\frac{1}{2}(J \star \phi + \phi \star J)$ is here to ensure that the simplification%
\footnote{
	In the case of a unimodular Lie group $\mathcal{G}$, the symmetrised  product correspond to the non-symmetrised one $\int J \star \phi$. This is explained in the trace cyclicity discussion of section \ref{subsec:trace_commu}.
}
\begin{align*}
	&- \frac{1}{2} \int \lHm(k)\ \phi(k) K(k) \phi(\ominus k) + \frac{1}{2} \int J \star \phi + \frac{1}{2} \int \phi \star J \\
	=  &- \frac{1}{2} \int \lHm(k)\ \phi(k) K(k) \phi(\ominus k) - \frac{1}{2} \int \lHm(k)\ J(k) K^{-1}(k) J(\ominus k)
\end{align*}
occurs, via the usual change of variable $\phi \to \phi + K^{-1}J$ in the first integral. Note that the previous equation only holds if
\begin{align}
	K^{-1}(\ominus k) = K^{-1}(k),
	\label{eq:prop_symm}
\end{align}
which is the case of our $K$.

Then, the perturbative expansion of the generating function writes
\begin{align}
\begin{aligned}
	\mathcal{Z}(J)
	&= \mathcal{Z}(0)\ \mathrm{exp}\left( - S_{\text{int}}\left(\frac{\partial}{\partial J} \right) \right) \mathrm{exp} \left( - \frac{1}{2} \int \lHm(k)\ J(k) K^{-1}(k) J(\ominus k) \right) \\
	&= \mathcal{Z}(0) \sum_{n=0}^\infty \frac{(-1)^n}{n!} S_{\text{int}}^n\left(\frac{\partial}{\partial J}\right) \mathrm{exp} \left( - \frac{1}{2} \int \lHm(k)\ J(k) K^{-1}(k) J(\ominus k) \right).
	\label{eq:gen_func_exp}
\end{aligned}
\end{align}

\paragraph{}
The $2$-point function at one-loop order corresponds to
\begin{align}
	\big\langle \phi(p) \phi(q) \big\rangle_{\text{1-loop}}
	= - \left. \frac{\partial}{\partial J(p)} \frac{\partial}{\partial J(q)}   S_{\text{int}} \left( \frac{\partial}{\partial J} \right) \mathrm{exp} \left( -  \frac{1}{2} \int \lHm(k)\ J(k) K^{-1}(k) J(\ominus k) \right) \right|_{J = 0}.
	\label{eq:2_point_1_loop}
\end{align}
where one-loop means that only the second term ($n=1$) in the exponential expansion of \eqref{eq:gen_func_exp} is considered. Note that we have $\frac{\partial J(p)}{\partial J(q)} = \delta(p \ominus q)$, so that 
\begin{align}
	\frac{\partial}{\partial J(p)} e^{-  \frac{1}{2} \int \lHm(k)\ J(k) K^{-1}(k) J(\ominus k)}
	&= - \frac{1 + \Delta(p)}{2} K^{-1}(p) J(\ominus p) \ e^{ -  \frac{1}{2} \int \lHm(k)\ J(k) K^{-1}(k) J(\ominus k)}.
	\label{eq:}
\end{align}
After computation and getting rid of disconnected components, one finds the relevant diagrams. Eight of these diagrams are planar, as pictured in Figure \ref{fig:planar}, and four are non-planar, see Figure \ref{fig:nonplanar}.

\begin{figure}[h]
    \begin{minipage}{.249\textwidth}
         \centering
    \begin{tikzpicture}[scale = 1.2]
        \draw[black] (-.1,-.1) rectangle (.1,.1);
        \draw[black] (-.7, -.7) node[anchor= east]{$\phi(q)$} to (-.1, -.1);
        \draw[black] (-.7,  .7) node[anchor= east]{$\phi(p)$} to (-.1,  .1);
        \draw[-{To}, black] (.1, .1) to (.6, .6) to[out= 45, in= 90] (.9,0)
            node[anchor = east]{$k$};
        \draw[black] (.1,-.1) to (.6,-.6) to[out=-45, in=-90] (.9,0);
        \draw (0,-.8) node[anchor = north]{$V\big((\ominus p) (\ominus q) (\ominus k)k \big)$} to (0,-.8);
    \end{tikzpicture}
    \end{minipage}%
    \begin{minipage}{.249\textwidth}
         \centering
    \begin{tikzpicture}[scale = 1.2]
        \draw[black] (-.1,-.1) rectangle (.1,.1);
        \draw[black] (.7,  .7) node[anchor= west]{$\phi(q)$} to (.1,  .1);
        \draw[black] (.7, -.7) node[anchor= west]{$\phi(p)$} to (.1, -.1);
        \draw[black] (-.1,-.1) to (-.6,-.6) to[out=-135, in=-90] (-.9,0);
        \draw[-{To}, black] (-.1, .1) to (-.6, .6) to[out= 135, in= 90] (-.9,0)
            node[anchor = west]{$k$};
        \draw (0,-.8) node[anchor = north]{$V\big((\ominus k)k (\ominus p) (\ominus q)\big)$} to (0,-.8);
    \end{tikzpicture}
    \end{minipage}%
    \begin{minipage}{.249\textwidth}
         \centering
    \begin{tikzpicture}[scale = 1.2]
        \draw[black] (-.1,-.1) rectangle (.1,.1);
        \draw[black] (-.7, -.7) node[anchor= east]{$\phi(p)$} to (-.1, -.1);
        \draw[black] ( .7, -.7) node[anchor= west]{$\phi(q)$} to ( .1, -.1);
        \draw[-{To}, black] (-.1,  .1) to (-.6,  .6) to[out=135, in=180] (0, .9)
            node[anchor = north]{$k$};
        \draw[black] (.1,.1) to (.6,.6) to[out=45, in=0] (0, .9);
        \draw (0,-.8) node[anchor = north]{$V\big((\ominus k) (\ominus p) (\ominus q)k\big)$} to (0,-.8);
    \end{tikzpicture}
    \end{minipage}%
    \begin{minipage}{.249\textwidth}
         \centering
    \begin{tikzpicture}[scale = 1.2]
        \draw[black] (-.1,-.1) rectangle (.1,.1);
        \draw[black] (-.7,.7) node[anchor= east]{$\phi(p)$} to (-.1,.1);
        \draw[black] ( .7,.7) node[anchor= west]{$\phi(q)$} to ( .1,.1);
        \draw[-{To}, black] (-.1, -.1) to (-.6, -.6) to[out=-135, in=180] (0, -.9)
            node[anchor = south]{$k$};
        \draw[black] (.1,-.1) to (.6,-.6) to[out=-45, in=0] (0, -.9);
        \draw (0,-.9) node[anchor = north]{$V\big((\ominus p)(\ominus k)k(\ominus q)\big)$} to (0,-.9);
    \end{tikzpicture}
    \end{minipage}
    
    \begin{minipage}{.249\textwidth}
         \centering
        \begin{tikzpicture}[scale = 1.2]
        \draw[black] (-.1,-.1) rectangle (.1,.1);
        \draw[black] (-.7, -.7) node[anchor= east]{$\phi(p)$} to (-.1, -.1);
        \draw[black] (-.7,  .7) node[anchor= east]{$\phi(q)$} to (-.1,  .1);
        \draw[-{To}, black] (.1, .1) to (.6, .6) to[out= 45, in= 90] (.9,0)
            node[anchor = east]{$k$};
        \draw[black] (.1,-.1) to (.6,-.6) to[out=-45, in=-90] (.9,0);
        \draw (0,-.8) node[anchor = north]{$V\big((\ominus q)(\ominus p)(\ominus k)k\big)$} to (0,-.8);
    \end{tikzpicture}
    \end{minipage}%
    \begin{minipage}{.249\textwidth}
         \centering
    \begin{tikzpicture}[scale = 1.2]
        \draw[black] (-.1,-.1) rectangle (.1,.1);
        \draw[black] (.7,  .7) node[anchor= west]{$\phi(p)$} to (.1,  .1);
        \draw[black] (.7, -.7) node[anchor= west]{$\phi(q)$} to (.1, -.1);
        \draw[black] (-.1,-.1) to (-.6,-.6) to[out=-135, in=-90] (-.9,0);
        \draw[-{To}, black] (-.1, .1) to (-.6, .6) to[out= 135, in= 90] (-.9,0)
            node[anchor = west]{$k$};
        \draw (0,-.8) node[anchor = north]{$V\big((\ominus k)k(\ominus q)(\ominus p)\big)$} to (0,-.8);
    \end{tikzpicture}
    \end{minipage}%
    \begin{minipage}{.249\textwidth}
         \centering
    \begin{tikzpicture}[scale = 1.2]
        \draw[black] (-.1,-.1) rectangle (.1,.1);
        \draw[black] (-.7, -.7) node[anchor= east]{$\phi(q)$} to (-.1, -.1);
        \draw[black] ( .7, -.7) node[anchor= west]{$\phi(p)$} to ( .1, -.1);
        \draw[-{To}, black] (-.1,  .1) to (-.6,  .6) to[out=135, in=180] (0, .9)
            node[anchor = north]{$k$};
        \draw[black] (.1,.1) to (.6,.6) to[out=45, in=0] (0, .9);
        \draw (0,-.8) node[anchor = north]{$V\big((\ominus k)(\ominus q)(\ominus p)k\big)$} to (0,-.8);
    \end{tikzpicture}
    \end{minipage}%
    \begin{minipage}{.249\textwidth}
         \centering
    \begin{tikzpicture}[scale = 1.2]
        \draw[black] (-.1,-.1) rectangle (.1,.1);
        \draw[black] (-.7,.7) node[anchor= east]{$\phi(q)$} to (-.1,.1);
        \draw[black] ( .7,.7) node[anchor= west]{$\phi(p)$} to ( .1,.1);
        \draw[-{To}, black] (-.1, -.1) to (-.6, -.6) to[out=-135, in=180] (0, -.9)
            node[anchor = south]{$k$};
        \draw[black] (.1,-.1) to (.6,-.6) to[out=-45, in=0] (0, -.9);
        \draw (0,-.9) node[anchor = north]{$V\big((\ominus q)(\ominus k)k(\ominus p)\big)$} to (0,-.9);
    \end{tikzpicture}
    \end{minipage}
    
    \caption{The four one-loop planar Feynman diagrams associated to the $2$-point function \eqref{eq:2_point_1_loop}.}
    \label{fig:planar}
\end{figure}

\begin{figure}[h]
    \begin{minipage}{.245\textwidth}
         \centering
    \begin{tikzpicture}[scale = 1.2]
        \draw[black] (-.1,-.1) rectangle (.1,.1);
        \draw[black] (-.7, -.7) node[anchor= east]{$\phi(p)$} to (-.1, -.1);
        \draw[black] ( .7,  .7) node[anchor= west]{$\phi(q)$} to ( .1,  .1);
        \draw[black,
            decoration={markings, mark=at position 0.75 with {\arrow{To}}},
            postaction={decorate}
        ] (-.1,  .1) to (-.6,  .6) to[out=135, in=135] (.45, .55);
        \node at (-.1,1.1) {$k$};
        \draw[black] ( .1, -.1) to ( .6, -.6) to[out=-45, in=-45] (.55, .45);
        \draw (0,-.8) node[anchor = north]{$V\big((\ominus k)(\ominus p)k(\ominus q)\big)$} to (0,-.8);
    \end{tikzpicture}
    \end{minipage}%
    \begin{minipage}{.245\textwidth}
         \centering
    \begin{tikzpicture}[scale = 1.2]
        \draw[black] (-.1,-.1) rectangle (.1,.1);
        \draw[black] (-.7, -.7) node[anchor= east]{$\phi(q)$} to (-.1, -.1);
        \draw[black] ( .7,  .7) node[anchor= west]{$\phi(p)$} to ( .1,  .1);
        \draw[black,
            decoration={markings, mark=at position 0.75 with {\arrow{To}}},
            postaction={decorate}
        ] (-.1,  .1) to (-.6,  .6) to[out=135, in=135] (.45, .55);
        \node at (-.1,1.1) {$k$};
        \draw[black] ( .1, -.1) to ( .6, -.6) to[out=-45, in=-45] (.55, .45);
        \draw (0,-.8) node[anchor = north]{$V\big((\ominus k)(\ominus q)k(\ominus p)\big)$} to (0,-.8);
    \end{tikzpicture}
    \end{minipage}%
    \begin{minipage}{.245\textwidth}
         \centering
    \begin{tikzpicture}[scale = 1.2]
        \draw[black] (-.1,-.1) rectangle (.1,.1);
        \draw[black] (-.7,  .7) node[anchor= east]{$\phi(p)$} to (-.1,  .1);
        \draw[black] ( .7, -.7) node[anchor= west]{$\phi(q)$} to ( .1, -.1);
        \draw[black,
            decoration={markings, mark=at position 0.75 with {\arrow{To}}},
            postaction={decorate}
        ] ( .1,  .1) to ( .6,  .6) to[out=  45, in=  45] (.55, -.45);
        \node at (1,0) {$k$};
        \draw[black] (-.1, -.1) to (-.6, -.6) to[out=-135, in=-135] (.45, -.55);
        \draw (0,-.8) node[anchor = north]{$V\big((\ominus p)(\ominus k)(\ominus q)k\big)$} to (0,-.8);
    \end{tikzpicture}
    \end{minipage}
    \begin{minipage}{.245\textwidth}
         \centering
    \begin{tikzpicture}[scale = 1.2]
        \draw[black] (-.1,-.1) rectangle (.1,.1);
        \draw[black] (-.7,  .7) node[anchor= east]{$\phi(q)$} to (-.1,  .1);
        \draw[black] ( .7, -.7) node[anchor= west]{$\phi(p)$} to ( .1, -.1);
        \draw[black,
            decoration={markings, mark=at position 0.75 with {\arrow{To}}},
            postaction={decorate}
        ] ( .1,  .1) to ( .6,  .6) to[out=  45, in=  45] (.55, -.45);
        \node at (1,0) {$k$};
        \draw[black] (-.1, -.1) to (-.6, -.6) to[out=-135, in=-135] (.45, -.55);
        \draw (0,-.8) node[anchor = north]{$V\big((\ominus q)(\ominus k)(\ominus p)k\big)$} to (0,-.8);
    \end{tikzpicture}
    \end{minipage}
    
    \caption{The four one-loop non-planar Feynman diagram associated to the $2$-point function \eqref{eq:2_point_1_loop}.}
    \label{fig:nonplanar}
\end{figure}

\paragraph{}
Using the vertex expression of \eqref{eq:action_mom}, one gets the planar vertices
\begin{subequations}
	\label{eq:vtx_plan}
\begin{align}
	V\big( (\ominus q) (\ominus p) (\ominus k) k \big)
	&= V\big( (\ominus k) k (\ominus q) (\ominus p) \big)
	= V\big( (\ominus q) (\ominus k) k (\ominus p) \big)
	= \delta(p \oplus q), \\
	V\big( (\ominus p) (\ominus q) (\ominus k) k \big)
	&= V\big( (\ominus k) k (\ominus p) (\ominus q) \big)
	= V\big( (\ominus p) (\ominus k) k (\ominus q) \big)
	= \delta(q \oplus p)
	= \Delta(q)\ \delta(p \oplus q)
\end{align}
\end{subequations}%
because $k$ directly cancels. Concerning $V\big( (\ominus k) (\ominus q) (\ominus p) k \big)$ and $V\big( (\ominus k) (\ominus p) (\ominus q) k \big)$, one uses the deformed cyclicity of the delta function \eqref{eq:delta_cycl} to have
\begin{subequations}
	\label{eq:vtx_plan_D}
\begin{align}
	V\big( (\ominus k) (\ominus q) (\ominus p) k \big)
	&= \Delta(k)^{-1}\ \delta(p \oplus q), \\
	V\big( (\ominus k) (\ominus p) (\ominus q) k \big)
	&= \Delta(k)^{-1}\ \delta(q \oplus p)
	= \Delta(q) \Delta(k)^{-1}\ \delta(p \oplus q).
\end{align}
\end{subequations}

In the same time the four non-planar contributions amounts to
\begin{subequations}
	\label{eq:vtx_nplan}
\begin{align}
	V \big( (\ominus k) (\ominus q) k (\ominus p) \big)
	&= \Delta(q)^{-1}\ V\big( (\ominus p) (\ominus k) (\ominus q) k \big)
	= \delta(p \ominus k \oplus q \oplus k), \\
	V \big( ( \ominus q) (\ominus k) (\ominus p) k \big)
	&= \Delta(q)^{-1}\ V \big( (\ominus k) (\ominus p) k (\ominus q) \big)
	= \Delta(k)^{-1}\ \delta(p \oplus k \oplus q \ominus k).	
\end{align}
\end{subequations}
To gather all this under one contribution, let us use that this vertex is integrated over $k$ so that, for any function $f$,
\begin{align*}
	\int \lHm(k)\ f(k)\ \Delta(k)^{-1}\ \delta(p \oplus k \oplus q \ominus k)
	&\overset{(k \to \ominus k)}{=} \int \lHm(k)\ f(\ominus k)\ \delta(p \ominus k \oplus q \oplus k).
\end{align*}
Regarding expressions \eqref{eq:vtx_plan}, \eqref{eq:vtx_plan_D} and \eqref{eq:vtx_nplan}, all planar diagrams are equal in the unimodular case. Correspondingly, all non-planar diagrams are also equal in this case.

\paragraph{}
Finally, the one-loop contribution writes
\begin{align}
\begin{aligned}
	\big\langle \phi(p) \phi(q) \big\rangle_{\text{1-loop}}
	= \frac{\lambda}{4!} \frac{1 + \Delta(q)}{2} 
	& \left( \delta(p \oplus q) \int \lHm(k)\ K^{-1}(k)\ \big(1 + \Delta(k)\big) \big(3 + \Delta(k)^{-1} \big) \right. \\
	&+ \left. \int \lHm(k)\ K^{-1}(k)\ \big(1 + \Delta(k) \big) \big( 1 + \Delta(k)^{-1} \big) \ \delta(p \ominus k \oplus q \oplus k) \right)
\end{aligned}
	\label{eq:2_point_nc}
\end{align}
where we used \eqref{eq:prop_symm}. Note that this expression simplifies to
 \begin{align}
\begin{aligned}
	\big\langle \phi(p) \phi(q) \big\rangle_{\text{1-loop}}
	= \frac{\lambda}{4!} \frac{1 + \Delta(q)}{2} 
	& \left( \delta(p \oplus q) \int \lHm(k)\ K^{-1}(k)\ \big(5 + 3 \Delta(k)\big) \right. \\
	&+ \left. \int \lHm(k)\ K^{-1}(k)\ \big(1 + \Delta(k) \big) \big( 1 + \Delta(k)^{-1} \big) \ \delta(p \ominus k \oplus q \oplus k) \right)
\end{aligned}
	\tag{\ref{eq:2_point_nc}}
\end{align}
by using \eqref{eq:int_prop_0}.

\subsection{Twisted trace and commutative limit}
\label{subsec:trace_commu}
\paragraph{}
Several comments can be made.

\paragraph{}
First, one should note that, contrary to a common misconception in noncommutative field theories, all the planar diagrams are not necessarily equal. In the same spirit, all the non-planar diagrams are not necessarily equal either. This is due to the presence of the modular function $\Delta$ in \eqref{eq:vtx_plan}, \eqref{eq:vtx_plan_D} and \eqref{eq:vtx_nplan}. Still, all the planar diagrams can be factorized to one contribution, and similarly for the non-planar diagrams, thus leading to only two terms in the $2$-point function \eqref{eq:2_point_nc}.

\paragraph{}
The commutative limit amounts to consider $\oplus = +$, $\ominus = -$, $\lHm(k) = \td^{d+1}k$ (the Lebesgue measure) and $\Delta = 1$, so that one recovers that all contributions \eqref{eq:vtx_plan}, \eqref{eq:vtx_plan_D} and \eqref{eq:vtx_nplan} are equal to $\delta(p + q)$.

Furthermore, the commutative limit of the two-point function \eqref{eq:2_point_nc} gives 
\begin{align}
	\big\langle \phi(p) \phi(q) \big\rangle_{\text{1-loop}}
	= \frac{\lambda}{2}\ \delta(p + q) \int \tdl{d+1}{k} K^{-1}(k),
	\label{eq:2_pt_commu}
\end{align}
which is the expected expression.

One can verify that if \eqref{eq:prop_div_UV} is fulfilled, then \eqref{eq:2_pt_commu} satisfies \ref{it:UV} and \ref{it:IR}. This underlines the necessity of considering \ref{it:fin} to define the UV/IR mixing.

\paragraph{}
A remark of another kind is in order. One can derive the (deformed) cyclicity property of the trace $\int \td^{d+1} x$ from the (deformed) cyclicity of the $\delta$ function \eqref{eq:delta_cycl}. That is
\begin{align}
\begin{aligned}
	\int \tdl{d+1}{x} f \star g (x)
	&= \int \lHm(p)\ \lHm(q)\ f(p) g(q)\ \delta(p \oplus q) \\
	&= \int \lHm(p)\ \lHm(q)\ \Delta(\ominus q)\ f(p) g(q)\ \delta(q \oplus p)
	= \int \tdl{d+1}{x} \Delta(g) \star f (x).
\end{aligned}
	\label{eq:trace_cycl}
\end{align}
where $\Delta$ in position space is now an operator one derives by inverse Fourier transform of $\Delta(\ominus\ \cdot)$. Therefore, the cyclicity property of the trace of $\algA$ depends only on the modular function of $\mathcal{G}$. If $\mathcal{G}$ is not unimodular, then the trace will be twisted and the twist corresponds to the inverse Fourier transform of the modular function. However, in the unimodular case, one concludes that the trace is cyclic.

\subsection{Divergences in the UV and the IR}
\label{subsec:UV_IR_div}
\paragraph{}
We now turn to the UV and IR analysis of the two point function. The following analysis takes its root in the fact that the expression \eqref{eq:2_point_nc} is valid for any propagator $K^{-1}$ satisfying \eqref{eq:prop_symm}.

\paragraph{}
Before beginning the analysis, let us write that for any $n \in \mathbb{Z}$,
\begin{align}
	\int \lHm(k)\ K^{-1}(k)\ \Delta(k)^n
	\overset{(k \to \ominus k)}{=}
	\int \lHm(k)\ K^{-1}(k)\ \Delta(k)^{-n-1},
	\label{eq:int_prop_n}
\end{align}
where we used that the propagator fulfils \eqref{eq:prop_symm}. This allows us to link positive and negative powers of $\Delta$ when integrating the propagator. Especially, one have that
\begin{subequations}
	\label{eq:int_prop}
\begin{align}
	\int \lHm(k)\ K^{-1}(k)
	&= 	\int \lHm(k)\ K^{-1}(k)\ \Delta(k)^{-1},
	\label{eq:int_prop_0}\\
	\int \lHm(k)\ K^{-1}(k)\ \Delta(k)
	&= 	\int \lHm(k)\ K^{-1}(k)\ \Delta(k)^{-2}. 
	\label{eq:int_prop_1}
\end{align}
\end{subequations}

\paragraph{}
Let us consider first the planar contribution. Since $\Delta >0$, one can write
\begin{align*}
	\int \lHm(k)\ K^{-1}(k)\ (5 + 3\Delta(k)) > 5 \int \lHm(k)\ K^{-1}(k),
\end{align*}
therefore, if $\int \lHm(k)\ K^{-1}(k)$ diverges in the UV, then the planar contribution is also diverging. This is what happened in most of the cases encountered in noncommutative scalar field theory, see for example \cite{Flik_1996, Minwalla_2000, Grosse_2006, Poulain_2019, Dimitrijevic_2018, Hersent_2023}, but also in the commutative case. Therefore, \ref{it:UV} is implied by
\begin{align}
	\int \lHm(k)\ K^{-1}(k) \text{ is diverging in the UV},
	\label{eq:prop_div_UV}
\end{align}
and we write \eqref{eq:prop_div_UV} $\Rightarrow$ \ref{it:UV}. In other words, the UV divergence is fully determined by the propagator.

\paragraph{}
Conversely, we consider that \eqref{eq:prop_div_UV} is not fulfilled, \textit{i.e.}\ the integral $\int \lHm(k)\ K^{-1}(k)$ is converging. Then, by continuity the $\Delta$ function is bounded, except if it diverges at $\pm \infty$. If it is bounded, then there exists some real number $M > 0$, that is $\Delta(k) < M$, thus
\begin{align*}
	\int \lHm(k)\ K^{-1}(k)\ (5 + 3\Delta(k))
	< (5 + 3M)\ \int \lHm(k)\ K^{-1}(k)
\end{align*}
which is finite. Otherwise, if the modular function diverges at infinity, then we use \eqref{eq:int_prop_1} to link it to its inverse power and write
\begin{align*}
	\int \lHm(k)\ K^{-1}(k)\ (5 + 3\Delta(k))
	= \int \lHm(k)\ K^{-1}(k)\ (5 + 3\Delta^{-2}(k)).
\end{align*}
Then, $\Delta^{-2}$ is continuous and bounded, and we can apply the previous reasoning. Thus, \eqref{eq:prop_div_UV} $\Leftrightarrow$ \ref{it:UV}.

There are examples of noncommutative settings where the quantity $\int \lHm(k)\ K^{-1}(k)$ is indeed finite. This may be the case on $\mathbb{R}^3_\lambda$ \cite{Fatollahi_2007, Komaie_Moghaddam_2007, Komaie_Moghaddam_2008, Vitale_2013}, but also on $\kappa$-Minkowski as in \eqref{eq:kM_plan}. 

	The fact that the mixing is present in \cite{Grosse_2006} and absent in \eqref{eq:kM_plan} for the $\kappa$-Minkowski space suggests that the choice of the noncommutative space does not ensure or prevent from having the mixing. Indeed, the previous result, together with the following analysis, suggest that the (integral of the) propagator of the theory is driving the mixing.

\paragraph{}
We now study the IR limit of the non-planar contribution. This limit corresponds to one of the two external momenta going to $0$, that is $p, q \to 0$. From there, one computes
\begin{align*}
	\delta(p \ominus k \oplus q \oplus k)
	&\underset{p \to 0}{\longrightarrow} \delta(k \oplus q \ominus k)
	= \Delta(k)\ \delta(q) \\
	\delta(p \ominus k \oplus q \oplus k)
	&\underset{q \to 0}{\longrightarrow} \delta(p)
\end{align*}
so that the non-planar contribution \eqref{eq:2_point_nc} writes
\begin{align*}
	\sim \int \lHm(k)\ K^{-1}(k)\ \Big(1 + \Delta(k)^{\pm 1} \Big)^2,
\end{align*}
the $\pm$ depending on weather $p$ or $q$ is sent to $0$. As $\Delta > 0$, one has
\begin{align*}
	\int \lHm(k)\ K^{-1}(k)\ \Big(1 + \Delta(k)^{\pm 1} \Big)^2
	> \int \lHm(k)\ K^{-1}(k)
\end{align*}
so that it diverges if \eqref{eq:prop_div_UV} is satisfied. Therefore, \eqref{eq:prop_div_UV} $\Rightarrow$ \ref{it:IR}.

Conversely, if $\int \lHm(k) K^{-1}(k)$ converges, then we can use \eqref{eq:int_prop} to have only positive or negative powers of $\Delta$ in the integral. Furthermore, as either $\Delta$ or $\Delta^{-1}$ is bounded, we can writes that it exists some $M > 0$ such that
\begin{align*}
	\int \lHm(k)\ K^{-1}(k)\ \Big(1 + \Delta(k)^{\pm 1} \Big)^2
	< M \int \lHm(k)\ K^{-1}(k)
\end{align*}
which converges. Therefore, \eqref{eq:prop_div_UV} $\Leftrightarrow$ \ref{it:IR}.

\paragraph{}
Thus, we showed that \eqref{eq:prop_div_UV} is a criteria for \ref{it:UV} and \ref{it:IR}. This is not quite surprising since this is already the case in the commutative setting, as one can see in \eqref{eq:2_pt_commu}. It seems rather that the noncommutative setting differ from the usual setting by the fact that \eqref{eq:prop_div_UV} can be avoided and/or that \ref{it:fin} is fulfilled.

\paragraph{}
Therefore, we need to study \ref{it:fin}. Indeed, a recurrent property of noncommutative field theory is that the non-planar contribution is UV finite for non-vanishing external momenta, even with a UV divergent planar contribution. Yet, asserting the UV finiteness of the non-planar contribution in the general setting, developed in this paper, seems non-trivial since $p \ominus k \oplus q \oplus k = 0$ has no simple solution for $k$ when $p, q \neq 0$. Therefore, one needs to look into this delta contribution.

The delta function in the interacting term of \eqref{eq:action_mom} is often called deformed conservation of momenta, since the deformed law $\oplus$ is present. Here, this deformed conservation law appears differently in the planar and non-planar diagrams. In the planar case \eqref{eq:vtx_plan} and \eqref{eq:vtx_plan_D}, the deformed conservation law gathers the ingoing and outgoing momenta, as one should expect. However, in the non-planar diagram \eqref{eq:vtx_nplan}, the deformed conservation of momenta also involves the internal momenta (here noted $k$).

The deformed conservation of momenta of the non-planar contribution $\delta(p \ominus k \oplus q \oplus k)$ thus mixes the external and internal momenta. This is a pure noncommutative feature since, in the commutative case, this $\delta$ function does not involves $k$. 

If one distinguish between pure noncommutative and pure commutative components of the momentum, that is 
\begin{align*}
	p_{||} \oplus q_{||} = p_{||} + q_{||}, &&
	p_\perp \oplus q_\perp \neq p_\perp + q_\perp,
\end{align*}
then one can write the non-planar contribution as
\begin{align*}
	\int \lHm(k)\ \tilde{K}^{-1}(k)\ \delta(p \ominus k \oplus q \oplus k)
	&= \delta(p_{||} + q_{||}) \int \lHm(k_\perp)\ \lHm(k_{||})\ \tilde{K}^{-1}(k_{||}, k_\perp)\ \delta(p_\perp \ominus k_\perp \oplus q_\perp \oplus k_\perp),
\end{align*}
where $\tilde{K}$ can be read from \eqref{eq:2_point_nc}. Since the $\oplus$ is non-trivial for the $\perp$ components, one can find a solution, called $k^*_\perp$ to the equation $p_\perp \ominus k_\perp \oplus q_\perp \oplus k_\perp = 0$, and so integrate over the delta. This writes
\begin{align}
	\int \lHm(k)\ \tilde{K}^{-1}(k)\ \delta(p \ominus k \oplus q \oplus k)
	&= \delta(p_{||} + q_{||}) \int \lHm(k_{||})\ \tilde{K}^{-1}(k_{||}, k^*_\perp),
	\label{eq:nplan_perp}
\end{align}
Therefore, no conservation rule for $p_{\perp}$ and $q_{\perp}$ holds anymore \textit{a priori}. The physical interpretation of this last statement, if it stands, needs to be deepened.

Going back to the study of \ref{it:fin}, \eqref{eq:nplan_perp} shows that \ref{it:fin} is equivalent to the convergence of $\int \lHm(k_{||})\ \tilde{K}^{-1}(k_{||}, k^*_\perp)$. This integral is non-trivial in the general case since $k^*_\perp$ can depend on $k_{||}$.

\paragraph{}
Some comments on the previous analysis are in order at this point.

\paragraph{}
One should note that the definition we took for the UV/IR mixing involves a UV divergence and a IR singularity that are mixed, but does not mean that the field theory is non-renormalisable. Indeed, if one considers the IR damping model \cite{Gurau_2009} in Moyal space, the two-point function at one-loop is both UV divergent and IR singular. The interest of the model relies in the fact that this two divergences are decoupled thanks to a counter term and so can be renormalised separately (for a comprehensive review see \cite{Blaschke_2010}). The study of the renormalisability of a theory verifying \ref{it:UV}, \ref{it:IR} and \ref{it:fin} is beyond the scope of this paper.

\paragraph{}
Furthermore, the present analysis can either allow to cope with the mixing, or to ensure its presence. Indeed, some physical models are based on the mixing, such as the exploration of IR contributions of UV origin in the context of the hierarchy problem \cite{Craig_2020}, or the emergent gravity theory of \cite{Steinacker_2007} from $U(n)$ gauge theory.

\section{Examples}
\label{sec:eg}
\paragraph{}
We here provide some concrete example of application of the previous model and their link to the existent computation of the UV/IR mixing. The first and pioneer example is the Moyal space (see section \ref{subsec:Moy}), where the mixing was first discovered \cite{Minwalla_2000}. Another well studied is the $\kappa$-Minkowski space (see section \ref{subsec:kM}) due to its relations with phenomenology (for a review on quantum gravity phenomenology see \cite{Addazi_2022}).

Note that this section is far from being exhaustive since we could apply the formalism developed in this paper to already studied noncommutative spaces, such as $\rho$-Minkowski \cite{Dimitrijevic_2018, Hersent_2023}.

\subsection{Moyal}
\label{subsec:Moy}
\paragraph{}
The Moyal space has a coordinate algebra given by
\begin{align}
	[x^\mu, x^\nu]_\star = i\Theta^{\mu\nu}
	\label{eq:Moy_coord_nc}
\end{align}
which is not of Lie algebra type \textit{a priori}. Still, by artificially considering the unit as a coordinate $x^{d+1} = 1$, one gets that
\begin{align}
	[x^\mu, x^\nu]_\star = i\Theta^{\mu\nu} x^{d+1}, &&
	[x^\mu, x^{d+1}]_\star = 0.
\end{align}
This relation is of Lie algebra type \eqref{eq:coord_nc} with $\tensor{f}{^{ab}_c} = i \Theta^{ab} \delta_c^{d+1} (1 - \delta^{a}_c) (1 - \delta^{b}_c)$, where $a,b,c = 0, \dots, d+1$. The corresponding group law is given by
\begin{subequations}
	\label{eq:Moy_Lie_gp}
\begin{align}
	p_\mu \oplus q_\mu &= p_\mu + q_\mu, &
	p_{d+1} \oplus q_{d+1} &= p_{d+1} + q_{d+1} + i p_\mu\Theta^{\mu\nu}q_\nu, \\
	\ominus p_\mu &= - p_\mu, &
	\ominus p_{d+1} &= - p_{d+1}
\end{align}
\end{subequations}
where $\mu = 0, \dots, d$. Since Moyal is a deformation of $\mathbb{R}^4$, the considered metric is the euclidean one $\mathrm{g}^{\mu\nu} = (+ \cdots +)$. Note that this approach was already taken in \cite{Flik_1996}. The $\oplus$ law for the $d+1^{\text{st}}$ component stands because $\det(\Theta) = 1$. The latter statement comes form the fact that $\Theta$ can be put under the form $\Theta = \mathrm{diag}(J, \dots, J)$, where $J = \begin{pmatrix} 0 & 1 \\ -1 & 0 \end{pmatrix}$. One can check that \eqref{eq:op_prop} are satisfied. Note that the antisymmetry of $\Theta$, which can be red directly from \eqref{eq:Moy_coord_nc}, is needed for \eqref{eq:op_prop_inv}. This group is unimodular so that $\Delta = 1$, and the Haar measure is the Lebesgue measure, \textit{i.e.}\ $\lHm(p) = \td^{d+1}p$.

\paragraph{}
Given an action of the form \eqref{eq:action}, all the planar diagrams are equal. Similarly, all the non-planar diagrams are also equal. This comes from the unimodularity of the group \eqref{eq:Moy_Lie_gp}. Therefore, we only have two contributions
\begin{subequations}
\begin{align}
	V(\text{planar})
	&= \delta(p + q), 
	\label{eq:Moy_vtx_plan}\\
	V(\text{non-planar})
	&= \delta(p_\mu + q_\mu)\ \delta(p_{d+1} + q_{d+1} + 2 p\Theta k),
	\label{eq:Moy_vtx_nplan}
\end{align}
\end{subequations}
where $p \Theta k = p_\mu \Theta^{\mu\nu} k_\nu$ and we used the antisymmetry of $\Theta$ through $p \Theta k = - k \Theta p$.

\paragraph{}
Therefore, the one-loop action is
\begin{align}
	\big\langle \phi(p) \phi(q) \big\rangle_{\text{1-loop}}
	= \frac{\lambda}{6}\ \delta(p + q)
	\int \tdl{d+1}{k} K^{-1}(k)\ \left(2 + e^{2 i p\Theta k} \right).
	\label{eq:Moy_2pt}
\end{align}
When considering the propagator $K(k) = \frac{1}{2} \big( k^\mu (\ominus k_\mu) + m^2 \big)$ one computes
\begin{align}
\begin{aligned}
	\big\langle \phi(p) \phi(q) \big\rangle_{\text{1-loop}}
	= \frac{2 \lambda}{3}\ \delta(p + q)
	\left( \int \tdl{}{\alpha} \left(\frac{\pi}{\alpha}\right)^{\frac{d+1}{2}} e^{-\alpha m^2} + \left( \frac{m}{p\Theta} \right)^{\frac{d-1}{2}} \mathcal{K}_{\frac{d-1}{2}}(2 m p\Theta) \right),
	\label{eq:Moy_2pt_comp}
\end{aligned}
\end{align}
where the left term is the divergent planar contribution, similar to the commutative contribution, and the right term is the non-planar contribution. We noted $\mathcal{K}$ the modified Bessel function of the second kind and recall that here $d$ is the space dimension. One can find the details of this computation in section \ref{apxsec:comp}.

As already depicted in the literature, the non-planar contribution is convergent for $p\Theta \neq 0$ and we have
\begin{align}
	2 \left( \frac{m}{p\Theta} \right)^{\frac{d-1}{2}} \mathcal{K}_{\frac{d-1}{2}}(2 m p\Theta)
	\underset{\substack{p\Theta \to 0 \\ m \to 0}}{\sim}
	\Gamma\left(\frac{d-1}{2}\right) \left( \frac{1}{(p\Theta)^2} \right)^{\frac{d-1}{2}}
	\label{eq:Moy_np_0}
\end{align}
thanks to \eqref{eq:apx_Bess0}. The latter limit is either when $p\Theta \to 0$ or when $m = 0$, therefore, the result is finite in the massless case when $p\Theta \neq 0$. Note that the limit $p\Theta \to 0$ either corresponds to the commutative limit $\Theta \to 0$ or to the IR limit $p \to 0$.

Therefore, from the previous analysis, the statement \eqref{eq:prop_div_UV} is fulfilled for this model on the Moyal space. Furthermore, we observe that the planar contribution diverges in the UV so that \ref{it:UV} is satisfied. The fact that this divergence is indeed UV need the regularization of the $2$-point function, which was already done in \cite{Minwalla_2000}. The non-planar contribution is IR singular with similar power then the planar UV divergence so that we also have \ref{it:IR}. Finally, the equation \eqref{eq:Moy_np_0} proves \ref{it:fin} in this case.

\subsection{\tops{$\kappa$}{kappa}-Minkowski}
\label{subsec:kM}
\paragraph{}
The $\kappa$-Minkowski space \cite{Lukierski_2017} has a coordinate algebra given by
\begin{align}
	[x^0, x^j]_\star = \frac{i}{\kappa} x^j, &&
	[x^j, x^k]_\star = 0,
\end{align}
where $j,k = 1, \dots, d$ are space indices, $\kappa$ is a constant with mass dimension. This amounts to taking $\tensor{f}{^{\mu\nu}_\rho} = \frac{i}{\kappa} (\delta^\mu_0 \delta^\nu_\rho - \delta^\nu_0 \delta^\mu_\rho)$ in \eqref{eq:coord_nc}. The corresponding group law is given by
\begin{subequations}
	\label{eq:mom_coord_kM_r}
\begin{align}
	p_0 \oplus q_0 &= p_0 + q_0, &
	p_j \oplus q_j &= p_j + e^{-p_0/\kappa} q_j \\
	\ominus p_0 &= -p_0, &
	\ominus p_j &= - e^{p_0/\kappa} p_j.
\end{align} 
\end{subequations}
Given these laws, one can check that \eqref{eq:op_prop} are satisfied. One can check that the right invariant Haar measure is the Lebesgue one $\rHm(p) = \td^{d+1}p$ and the left invariant Haar measure is $\lHm(p) = e^{dp_0/\kappa} \td^{d+1}p$, where $d$ is the space dimension. Therefore, this group is not unimodular and has a modular function given by $\Delta(p) = e^{dp_0/\kappa}$. For a more detailed computation see for example \cite{Poulain_2018}.

\paragraph{}
Given an action of the form \eqref{eq:action}, one computes the one-loop $2$-point function thanks to \eqref{eq:2_point_nc},
\begin{align}
\begin{aligned}
	\big\langle \phi(p) \phi(q) \big\rangle_{\text{1-loop}}
	&= \frac{\lambda}{4!} \frac{1+ e^{dq_0/\kappa}}{2}\ \delta(p \oplus q) \int \tdl{d+1}{k} K^{-1}(k)\ e^{dk_0/\kappa} \big(5 + 3 e^{dk_0/\kappa} \big) \\
	&+ \frac{\lambda}{4!} \frac{1+ e^{dq_0/\kappa}}{2}\ \delta(p_0 + q_0) \int \tdl{d+1}{k} K^{-1}(k)\ \Big(1 + e^{dk_0/\kappa} \Big)^2 \\
	&\times e^{-d k_0/\kappa} \delta\!\left( e^{-k_0/\kappa} p_j + e^{- p_0)/\kappa} q_j + (1 - e^{-p_0 /\kappa}) k_j \right)
\end{aligned}
	\label{eq:2_point_kM}
\end{align}

\paragraph{}
We now go to the computation of the two point function with the propagator of \eqref{eq:action_mom}. We take the metric to be the Minkowski one $\mathrm{g}^{\mu\nu} = (+ - \cdots -)$. Let us compute first the planar contribution
\begin{align}
\begin{aligned}
	\int \tdl{d+1}{k} K^{-1}(k)\ e^{dk_0/\kappa} \big(5 + 3e^{dk_0/\kappa} \big)
	&= 4 \pi \left( \frac{4\pi \kappa m}{d} \right)^{\frac{d-1}{2}} \left( 5\ \mathcal{K}_{\frac{d-1}{2}} \left( \frac{m d}{2 \kappa} \right) + 3^{\frac{d+1}{2}} \mathcal{K}_{\frac{d-1}{2}} \left( \frac{3 m d}{2 \kappa} \right) \right)\\
	& \!\!\!\! \underset{\substack{\kappa \to +\infty \\ m \to 0}}{\sim} 2 \pi\ \Gamma\left(\frac{d-1}{2}\right) \left( \frac{4^2 \pi \kappa^2}{d^2} \right)^{\frac{d-1}{2}} \left( 5 + 3^{d} \right),
\end{aligned}
	\label{eq:kM_plan}
\end{align}
where $\mathcal{K}$ is the modified Bessel function of the second kind and recall that here $d$ is the space dimension. The last equivalent is taken either when $\kappa \to + \infty$, corresponding to the commutative limit, or when $m = 0$ which is the massless case. The details of this computation is gathered in appendix \ref{apxsec:comp}. 

This integral converges for a finite $\kappa$, even for $m=0$. $\kappa$ is here playing the role of the UV cut-off, a behaviour Snyder \cite{Snyder_1947} was hoping for. Therefore, in $d+1 = 4$ dimensions this integral diverges as $\kappa^2$.

\paragraph{}
Now we turn to the non-planar contribution. Considering the IR limit, one obtains
\begin{align}
\begin{aligned}
	 \delta\!\left( e^{-k_0/\kappa} p_j + e^{- p_0)/\kappa} q_j + (1 - e^{-p_0 /\kappa}) k_j \right)
	&\underset{p \to 0}{\longrightarrow} \delta (q_j), \\
	 \delta\!\left( e^{-k_0/\kappa} p_j + e^{- p_0)/\kappa} q_j + (1 - e^{-p_0 /\kappa}) k_j \right)
	&\underset{q \to 0}{\longrightarrow} e^{- d k_0 / \kappa} \delta (p_j).
\end{aligned}
	\label{eq:kM_delta_IR}
\end{align}

This leads to
\begin{align}
\begin{aligned}
	\int \tdl{d+1}{k} & K^{-1}(k)\ e^{\left(\frac{1}{2} \pm \frac{1}{2} \right) d k_0/\kappa} \left( 1 + e^{dk_0/\kappa} \right)^2 \\
	&= 4 \pi \left( \frac{2\pi \kappa m}{d} \right)^{\frac{d-1}{2}} \left( 
	\left(\pm \frac{1}{2} \right)^{\frac{1-d}{2}} \mathcal{K}_{\frac{d-1}{2}} \left( \frac{m d}{\kappa} \left(\pm \frac{1}{2} \right) \right) 
	+ 2 \left(1 \pm \frac{1}{2} \right)^{\frac{1-d}{2}} \mathcal{K}_{\frac{d-1}{2}} \left( \frac{m d}{\kappa} \left(1 \pm \frac{1}{2} \right) \right) \right. \\
	& + \left. \left(2 \pm \frac{1}{2} \right)^{\frac{1-d}{2}} \mathcal{K}_{\frac{d-1}{2}} \left( \frac{m d}{\kappa} \left(u + \frac{3}{2} \right) \right) \right) \\
	& \!\!\!\! \underset{\substack{\kappa \to +\infty \\ m \to 0}}{\sim} 2 \pi\ \Gamma\left(\frac{d-1}{2}\right) \left( \frac{4 \pi \kappa^2}{d^2} \right)^{\frac{d-1}{2}} \left( \left(\pm \frac{1}{2}\right)^{1-d} + 2 \left(1 \pm \frac{1}{2}\right)^{1-d} + \left(2 \pm \frac{1}{2}\right)^{1-d}\right),
\end{aligned}
	\label{eq:kM_nplan_IR}
\end{align}
where $\pm$ depends on whether $p$ or $q$ is going to $0$. The details of this computation can be found in appendix \ref{apxsec:comp}. Therefore, the non-planar computation is finite in the IR, for a finite $\kappa$. Here, $\kappa$ plays also the role of a cut-off and the integral diverges as $\kappa^2$ in $d+1 = 4$ dimensions.

Finally, for a finite $\kappa$, the computation \eqref{eq:kM_plan} shows that \eqref{eq:prop_div_UV} and \ref{it:UV} are not fulfilled. Furthermore, \ref{it:IR} does not stand in view of computation \eqref{eq:kM_nplan_IR}.

\paragraph{}
We now turn to the computation of the non-planar contribution in the case where the external momenta are not zero, \textit{i.e.}\ $p, q \neq 0$ (not light-like). If we further consider $p$ to be time-like, then $p_0^2 >  p_j^2 \geqslant 0$, so that $p_0 \neq 0$. Thus, the delta in \eqref{eq:2_point_kM} gives
\begin{align*}
	\delta\!\left( e^{-k_0/\kappa} p_j + e^{- p_0)/\kappa} q_j + (1 - e^{-p_0 /\kappa}) k_j \right)
	= \frac{1}{\left| 1 - e^{-p_0/\kappa} \right|^d}\ \delta\!\left(k_j - \frac{ e^{-k_0/\kappa} p_j + e^{-p_0/\kappa} q_j}{1 - e^{-p_0/\kappa}}\right),
\end{align*}
so that
\begin{align}
\begin{aligned}
	\int \tdl{d+1}{k} K^{-1}(k) &\ \Big(1 + e^{dk_0/\kappa} \Big)^2 \delta\!\left( e^{-k_0/\kappa} p_j + e^{- p_0/\kappa} q_j + (1 - e^{-p_0 /\kappa}) k_j \right) \\
	&= 2 \int \tdl{}{k_0} \frac{\Big(1 + e^{idk_0/\kappa} \Big)^2}{ \left| 1 - e^{-p_0/\kappa} \right|^d \left( k_0^2 + e^{ik_0/\kappa} \left( \frac{e^{-i k_0/\kappa} p_j + e^{-p_0/\kappa} q_j}{1 - e^{-p_0/\kappa}} \right)^2 + m^2 \right)}.
\end{aligned}
	\label{eq:kM_nplan}
\end{align}
The detailed computation can be found in appendix \ref{apxsec:comp}. As $p_0 \neq 0$, this integral is convergent, since it converges absolutly. Indeed, the denominator does not vanish, even when $k_0 = 0$, and at infinities the integrand behaves (in absolute value) as $k_0^{-2}$. Thus, the requirement \ref{it:fin} remains satisfied. The explicit computation of the integral could not be performed by the author though.

As already mentioned in section \ref{subsec:UV_IR_div}, this integral has \textit{a priori} no momentum conservation delta function for the pure noncommutative component. To make the link with the latter section, one has here $|| = \{0\}$ and $\perp = \{1, \dots, d\}$. Note that the disappearance of the conservation of momenta for the spatial components was already observed in \cite{Kosinski_2000, Amelino_Camelia_2002, Grosse_2006} for the case of $\kappa$-Minkowski.

\subsubsection*{Momentum coordinates invariance}
\paragraph{}
From there, we want to compute the $2$-point function in another momentum coordinate system. Above, we chose the momentum coordinate \eqref{eq:mom_coord_kM_r} associated to the time-to-the-right ordering of the exponential, like the second expression of \eqref{eq:mom_order}. We now choose the in-exponential-sum ordering, corresponding to the first expression of \eqref{eq:mom_order}. This gives
\begin{align}
	p_\mu \oplus q_\mu &= \frac{p_0 + q_0}{e^{-q_0/\kappa} - e^{p_0/\kappa}} \left( \frac{1 - e^{p_0/\kappa}}{p_0} p_\mu + \frac{e^{-q_0/\kappa} - 1}{q_0} q_\mu \right) , &
	\ominus p_\mu &= -p_\mu,
	\label{eq:mom_coord_kM_s}
\end{align}
where $\mu = 0, \dots, d$. Note that the previous expression is finite when $p_0, q_0 \to 0$ and that $p_0 \oplus q_0 = p_0 + q_0$. Furthermore, one can link both coordinates expressions \eqref{eq:mom_coord_kM_r} and \eqref{eq:mom_coord_kM_s} through
\begin{align}
	p_j \oplus_{\text{right}} q_j 
	&= \frac{1}{g\left( \frac{p_0 + q_0}{\kappa} \right)} \left( g\left(\frac{p_0}{\kappa}\right) p_j \oplus_{\text{sum}} g\left(\frac{q_0}{\kappa}\right) q_j \right),
\end{align}
where $g(x) = \frac{x}{1 - e^{-x}}$. Note that this relation is not a simple rescaling since the prefactor $g\left(\frac{p_0 + q_0}{\kappa}\right)$ depends both on $p$ and $q$.

\paragraph{}
This group is also not unimodular, but since the time component of the coordinates is unchanged, the modular function is still $\Delta(p) = e^{d p_0 / \kappa}$. However, the left invariant Haar measure changed to $\lHm(p) = \left(\frac{1 - e^{p_0 / \kappa}}{p_0}\right)^d \tdl{d+1}{p}$.

\paragraph{}
Given an action of the form \eqref{eq:action}, one computes the one-loop $2$-point function thanks to \eqref{eq:2_point_nc},
\begin{align}
\begin{aligned}
	\big\langle \phi(p) \phi(q) \big\rangle_{\text{1-loop}}
	&= \frac{\lambda}{4!} \frac{1 + e^{dq_0/\kappa}}{2}\ \delta(p \oplus q)\ \int \tdl{d+1}{k} K^{-1}(k)\ \left(\frac{1 - e^{k_0 / \kappa}}{k_0}\right)^d (5 + 3e^{d k_0/\kappa}) \\
	&+ \frac{\lambda}{4!} \frac{1 + e^{dq_0/\kappa}}{2}\ \delta(p_0 + q_0) \int \tdl{d+1}{k} K^{-1}(k)\ \left(\frac{1 - e^{k_0 / \kappa}}{k_0}\right)^d \Big(1 + e^{-dk_0/\kappa} \Big) \Big(1 + e^{dk_0/\kappa} \Big) \\
	&\times \delta\!\left( \frac{1 - e^{p_0 / \kappa}}{p_0} \big(p_j + e^{k_0 / \kappa} q_j \big) - (1 + e^{p_0/\kappa})  \frac{e^{k_0 / \kappa} - 1}{k_0} k_j \right)
\end{aligned}
	\label{eq:2_point_kM_s}
\end{align}

One can show that this $2$-point function is finite, after Wick rotation, for the propagator \eqref{eq:action_mom} and the Minkowski metric $\mathrm{g}^{\mu\nu} = (+ - \cdots -)$. The explicit expression of the $2$-point function was not found by the author.

\paragraph{}
However, the primary handling of the integrals of \eqref{eq:2_point_kM_s} suggests that the $2$-point function might give different results in the two coordinate systems. This can be simply highlighted by the planar contribution, since the quantity $\int \lHm(p)\ K^{-1}(p)$ is coordinate dependant. Explicitly, one has
\begin{subequations}
	\label{eq:prop_kM}
\begin{align}
	\int \lHm(k)\ K^{-1}(k)
	&= \int \tdl{d+1}{k} \frac{e^{dk_0/2\kappa}}{-k_0^2 + k_j^2 + m^2},
	\label{eq:prop_kM_r} \\
	\int \lHm(k)\ K^{-1}(k)
	&= \int \tdl{d+1}{k} \left( - \frac{\sinh\left( \frac{k_0}{2\kappa} \right)}{k_0} \right)^d \frac{e^{dk_0/2\kappa}}{-k_0^2 + k_j^2 + m^2},
	\label{eq:prop_kM_s}
\end{align}
\end{subequations}
where \eqref{eq:prop_kM_r} corresponds to the time-to-the-right ordering and \eqref{eq:prop_kM_s} to the in-exponential-sum one. Note that \eqref{eq:prop_kM_r} was obtained after a change of variable $k_j \to e^{-k_0 / 2\kappa} k_j$ for all $j = 1, \dots, d$. Nevertheless, the coordinate dependence of the $2$-point function only makes sense to the full loop computation which is beyond the scope of this paper.

\paragraph{}
Should the $2$-point function be coordinate dependent, one would need at least to ensure that the measurable quantities are coordinate independent. Still, some authors \cite{2_point} claim that a $2$-point function is an observable. In this case, the analysis of this paper would need to be made ordering independent.

\section{Conclusion}
\paragraph{}
We studied in this paper the UV/IR mixing mechanism of a generic noncommutative $\phi^4$ theory for Lie algebra type noncommutativity \eqref{eq:coord_nc}. We tried to define unambiguously the UV/IR mixing through the requirement \ref{it:UV}, \ref{it:IR} and \ref{it:fin} and capture a simple criterion that would be equivalent to the former requirements. It was shown in section \ref{subsec:UV_IR_div} that the divergence of the integral of the propagator \eqref{eq:prop_div_UV} is equivalent to \ref{it:UV} and \ref{it:IR}. However, the study of \ref{it:fin} being non-trivial, no link between \eqref{eq:prop_div_UV} and \ref{it:fin} was found. The closer we could get was \eqref{eq:nplan_perp}. Since the requirement \ref{it:fin} is essential to distinguish the noncommutative case from the commutative one, this point needs to be further explored.

This momentum based formalism was applied to the Moyal and the $\kappa$-Minkowski spaces. Usual results on Moyal were recovered and new ones were explored for $\kappa$-Minkowski. The requirements \ref{it:UV} to \ref{it:fin} and their relation to the criterion \eqref{eq:prop_div_UV} were exemplified on these two spaces. The considered theory on $\kappa$-Minkowski was shown to be divergence free.

\paragraph{}
Apart from this, two side results were obtained. The first one is the relation between the cyclicity of the integral and the modular function of the group of momenta given by equation \eqref{eq:trace_cycl}. Note that the non-cyclicity of the integral was considered to be the main difficulty in building a gauge invariant Yang-Mills like action on $\kappa$-Minkowski \cite{Hersent_2022}. It could hint for a systematic method to restore cyclicity or deal with non-cyclicity in other ways \cite{Mathieu_2020}.

Furthermore, we studied the ordering problem \eqref{eq:mom_order}, which is specific to the noncommutative case. Indeed, different orderings can trigger different laws for momentum (deformed) addition and so different results of the $2$-point function. In the case of $\kappa$-Minkowski, the change of ordering amounts to a change of coordinate of momentum. Therefore, we explored two computations of the $2$-point function in $\kappa$-Minkowski with two coordinate choices. However, due to cumbersome expressions \eqref{eq:prop_kM} the computation could not be carried out. Close but different expressions \eqref{eq:prop_kM} suggest that the results are different, coming from the fact that the action \eqref{eq:action_mom} is not diffeomorphic invariant and so ordering invariant. The latter hand waving arguments needs to be further studied.

%\newpage
\section*{Acknowledgement}
\paragraph{}
The author thanks M.\ Arzano, M.\ Dimitrijevi\'{c}-\'{C}iri\'{c}, S.\ Franchino-Vi\~{n}as, L.\ Ferdinand, J.\ Kowalski-Glikman, F.\ Lizzi, P.\ Martinetti, G.\ Nieuviarts, H.\ Steinacker, P.\ Vitale and J.-C.\ Wallet for fruitful discussions.

The author thanks S.\ S.\ Jabbari and A.\ H.\ Fatollahi for reading suggestions.

The author is also grateful to S.\ Koren for interests and enlightening comments, especially the necessity of \ref{it:fin} in the UV/IR mixing definition.

The author finally thanks the Action CA18108 QG-MM, ``Quantum Gravity Phenomenology in the multi-messengers approach'', and the Action CA21109 CaLISTA, ``Cartan geometry, Lie, Integrable Systems, quantum group Theories for Applications'', from the European Cooperation in Science and Technology (COST).

\appendix

\section{Formula}
\label{apxsec:form}
\paragraph{}
This section aims at gathering useful formulas for the above computation and so without proof.

\paragraph{}
First, the one-dimensional Gaussian integral writes
\begin{align}
	\int_{-\infty}^{+\infty} \tdl{}{x} e^{-ax^2 + bx+ c}
	&= \sqrt{\frac{\pi}{a}}\ e^{\frac{b^2}{4a} + c}
	\label{eq:apx_gauss}
\end{align}
where $a > 0$.

\paragraph{}
We will also need
\begin{align}
	\int_0^{+\infty} \tdl{}{x} \frac{1}{x^n} e^{-ax-\frac{b}{x}}
	&= 2 \left(\frac{b}{a}\right)^{\frac{1-n}{2}} \mathcal{K}_{n-1} \Big(2\sqrt{ab} \Big)
	\label{eq:apx_Bess}
\end{align}
for $a, b, n > 0$ and where $\mathcal{K}$ is the modified Bessel function of the second kind. The behaviour around $0$ of this latter function, for $\alpha > 0$, is given by
\begin{align}
	\mathcal{K}_{\alpha}(x)
	\underset{x \to 0}{\sim}
	\frac{\Gamma(\alpha)}{2} \left( \frac{2}{x} \right)^\alpha
	\label{eq:apx_Bess0}
\end{align}
where $\Gamma$ is the Euler gamma function.

\section{Computations}
\label{apxsec:comp}
\paragraph{}
In this section, we gather some details of previous computations.

\subsection{Moyal}
\begin{align}
\begin{aligned}
	\int \tdl{d+1}{k} K^{-1}(k)\ \left(2 + e^{2 i p\Theta k} \right)
	&= 2 \int \tdl{d+1}{k} \frac{2 + e^{2 i p\Theta k}}{k^2 + m^2} \\
	&= 2 \int \tdl{d+1}{k} \int_0^{+\infty} \tdl{}{\alpha} \left(2 + e^{2 i p\Theta k} \right) e^{-\alpha k^2 - \alpha m^2} \\
	& \!\! \overset{\eqref{eq:apx_gauss}}{=} 2 \int \tdl{}{\alpha} \left(\frac{\pi}{a}\right)^{\frac{d+1}{2}} \left( 2 + e^{\frac{- (p\Theta)^2}{\alpha}} \right) e^{-\alpha m^2} \\
	& \!\! \overset{\eqref{eq:apx_Bess}}{=} 4 \left( \int \tdl{}{\alpha} \left(\frac{\pi}{a}\right)^{\frac{d+1}{2}} e^{-\alpha m^2} + \left( \frac{m}{p\Theta} \right)^{\frac{d-1}{2}} \mathcal{K}_{\frac{d-1}{2}}(2 m p\Theta) \right)
\end{aligned}
	\tag{\ref{eq:Moy_2pt_comp}}
\end{align}
where we noted $k^2 = k_\mu k^\mu = k_0^2 + \cdots + k_d^2$.

\subsection{\tops{$\kappa$}{kappa}-Minkowski}
\begin{align}
\begin{aligned}
	\int \tdl{d+1}{k} K^{-1}(k)\ &e^{dk_0/\kappa} \big(5 + 3 e^{dk_0/\kappa} \big)
	= -2i \int \tdl{d+1}{k} \frac{5 e^{dk_0/\kappa} + 3 e^{2dk_0/\kappa}}{- k_0^2 + e^{k_0/\kappa}k_j^2 + m^2} \\
	&= -2i \int \tdl{d+1}{k} \int_0^{+\infty} \tdl{}{\alpha} (5e^{dk_0/\kappa} + 3e^{2dk_0/\kappa}) e^{-\alpha(-k_0^2 + e^{k_0/\kappa}k_j^2 + m^2)} \\
	& \!\! \overset{\eqref{eq:apx_gauss}}{=} -2i \int \tdl{}{k_0} \int \tdl{}{\alpha} (5e^{dk_0/\kappa} + 3e^{2dk_0/\kappa}) \left( \frac{\pi}{\alpha e^{k_0/\kappa}} \right)^{d/2} e^{\alpha k_0^2 - \alpha m^2} \\
	& \!\!\!\! \overset{\text{(Wick)}}{=} 2 \int \tdl{}{k_0} \int \tdl{}{\alpha} (5e^{idk_0/\kappa} + 3e^{2idk_0/\kappa}) \left( \frac{\pi}{\alpha}\right)^{d/2} e^{-idk_0/2\kappa} e^{-\alpha k_0^2 - \alpha m^2} \\
	&= 2 \int \tdl{}{k_0} \int \tdl{}{\alpha} \left( \frac{\pi}{\alpha}\right)^{\frac{d}{2}} \left( 5 e^{-\alpha k_0^2 +idk_0/2\kappa - \alpha m^2} + 3 e^{-\alpha k_0^2 + 3idk_0/2\kappa - \alpha m^2} \right) \\
	& \!\! \overset{\eqref{eq:apx_gauss}}{=} 2 \int \tdl{}{\alpha} \left( \frac{\pi}{\alpha}\right)^{\frac{d+1}{2}} \left(5\ e^{-\alpha m^2 - \frac{1}{\alpha} \left(\frac{d}{4\kappa}\right)^2} + 3\ e^{-\alpha m^2 - \frac{1}{\alpha} \left(\frac{3d}{4\kappa}\right)^2} \right) \\
	& \!\! \overset{\eqref{eq:apx_Bess}}{=} 4 \pi \left( \frac{4\pi \kappa m}{d} \right)^{\frac{d-1}{2}} \left( 5\ \mathcal{K}_{\frac{d-1}{2}} \left( \frac{m d}{2 \kappa} \right) + 3^{\frac{d+1}{2}} \mathcal{K}_{\frac{d-1}{2}} \left( \frac{3 m d}{2 \kappa} \right) \right)\\
	& \!\!\!\! \underset{\substack{\kappa \to +\infty \\ m \to 0}}{\sim} 2 \pi\ \Gamma\left(\frac{d-1}{2}\right) \left( \frac{4^2 \pi \kappa^2}{d^2} \right)^{\frac{d-1}{2}} \left( 5 + 3^{d} \right),
\end{aligned}
	\tag{\ref{eq:kM_plan}}
\end{align}
where the last line is obtained thanks to \eqref{eq:apx_Bess0}, considering $d > 1$.

\begin{align}
\begin{aligned}
	\int \tdl{d+1}{k} & K^{-1}(k)\ e^{u d k_0/\kappa} \left( 1 + e^{dk_0/\kappa} \right)^2
	= -2i \int \tdl{d+1}{k} e^{u d k_0/\kappa} \frac{\left( 1 + e^{dk_0/\kappa} \right)^2}{- k_0^2 + e^{k_0/\kappa}k_j^2 + m^2} \\
	&= -2i \int \tdl{d+1}{k} \int_0^{+\infty} \tdl{}{\alpha} e^{u d k_0/\kappa} \left( 1 + e^{dk_0/\kappa} \right)^2 e^{-\alpha(-k_0^2 + e^{k_0/\kappa}k_j^2 + m^2)} \\
	& \!\! \overset{\eqref{eq:apx_gauss}}{=} -2i \int \tdl{}{k_0} \int \tdl{}{\alpha} \left( \frac{\pi}{\alpha} \right)^{d/2} e^{\left(u - \frac{1}{2}\right) d k_0/\kappa} \left( 1 + e^{dk_0/\kappa} \right)^2 e^{\alpha k_0^2 - \alpha m^2} \\
	& \!\!\!\! \overset{\text{(Wick)}}{=} 2 \int \tdl{}{k_0} \int \tdl{}{\alpha} \left( \frac{\pi}{\alpha} \right)^{d/2} e^{\left(u - \frac{1}{2}\right) i d k_0/\kappa} \left( 1 + e^{idk_0/\kappa} \right)^2 e^{-\alpha k_0^2 - \alpha m^2} \\
	& \!\! \overset{\eqref{eq:apx_gauss}}{=} 2 \int \tdl{}{\alpha} \left( \frac{\pi}{\alpha}\right)^{\frac{d+1}{2}} e^{-\alpha m^2} \left(\ 
	e^{- \frac{1}{\alpha} \left(\frac{d \left(u - \frac{1}{2}\right)}{2\kappa}\right)^2} 
	+ 2 e^{- \frac{1}{\alpha} \left(\frac{d \left(u + \frac{1}{2}\right)}{2\kappa}\right)^2}
	+ e^{- \frac{1}{\alpha} \left(\frac{d \left(u + \frac{3}{2}\right)}{2\kappa}\right)^2} \right) \\
	& \!\! \overset{\eqref{eq:apx_Bess}}{=} 4 \pi \left( \frac{2\pi \kappa m}{d} \right)^{\frac{d-1}{2}} \left( 
	\left(u - \frac{1}{2} \right)^{\frac{1-d}{2}} \mathcal{K}_{\frac{d-1}{2}} \left( \frac{m d}{\kappa} \left(u - \frac{1}{2} \right) \right) 
	+ 2 \left(u + \frac{1}{2} \right)^{\frac{1-d}{2}} \mathcal{K}_{\frac{d-1}{2}} \left( \frac{m d}{\kappa} \left(u + \frac{1}{2} \right) \right) \right. \\
	& + \left. \left(u + \frac{3}{2} \right)^{\frac{1-d}{2}} \mathcal{K}_{\frac{d-1}{2}} \left( \frac{m d}{\kappa} \left(u + \frac{3}{2} \right) \right) \right)\\
	& \!\!\!\! \underset{\substack{\kappa \to +\infty \\ m \to 0}}{\sim} 2 \pi\ \Gamma\left(\frac{d-1}{2}\right) \left( \frac{4 \pi \kappa^2}{d^2} \right)^{\frac{d-1}{2}} \left( \left(u - \frac{1}{2}\right)^{1-d} + 2 \left(u + \frac{1}{2}\right)^{1-d} + \left(u + \frac{3}{2}\right)^{1-d}\right),
\end{aligned}
	\tag{\ref{eq:kM_nplan_IR}}
\end{align}
where $u = 0, 1$ depending whether $p \to 0$ or $q \to 0$ as one can read from \eqref{eq:kM_delta_IR}. The last line is obtained thanks to \eqref{eq:apx_Bess0}, considering $d > 1$.

\begin{align}
\begin{aligned}
	\int \tdl{d+1}{k} K^{-1}(k) &\ \Big(1 + e^{dk_0/\kappa} \Big)^2 \delta\!\left( e^{-k_0/\kappa} p_j + e^{- p_0/\kappa} q_j + (1 - e^{-p_0 /\kappa}) k_j \right) \\
	&= -2i \int \tdl{d+1}{k} \frac{\Big(1 + e^{dk_0/\kappa} \Big)^2}{\left| 1 - e^{-p_0/\kappa} \right|^d \left(-k_0^2 + e^{k_0/\kappa} k_j^2 + m^2\right)}\ \delta\!\left(k_j - \frac{e^{-k_0/\kappa} p_j + e^{- p_0/\kappa} q_j}{1 - e^{-p_0/\kappa}} \right) \\
	&= -2i \int \tdl{}{k_0} \frac{\Big(1 + e^{dk_0/\kappa} \Big)^2}{ \left| 1 - e^{-p_0/\kappa} \right|^d \left(-k_0^2 + e^{k_0/\kappa} \left( \frac{e^{-k_0/\kappa} p_j + e^{- p_0/\kappa} q_j}{1 - e^{-p_0/\kappa}} \right)^2 + m^2 \right)} \\
	& \!\!\!\! \overset{\text{(Wick)}}{=} 2 \int \tdl{}{k_0} \frac{\Big(1 + e^{idk_0/\kappa} \Big)^2}{ \left| 1 - e^{-p_0/\kappa} \right|^d \left( k_0^2 + e^{ik_0/\kappa} \left( \frac{e^{-i k_0/\kappa} p_j + e^{-p_0/\kappa} q_j}{1 - e^{-p_0/\kappa}} \right)^2 + m^2 \right)}
\end{aligned}
	\tag{\ref{eq:kM_nplan}}
\end{align}

\end{document}